\documentclass{article}

\usepackage{iclr2027_conference,times}
\usepackage{microtype}
\usepackage{amsmath,amssymb}
\usepackage{booktabs}
\usepackage{array}
\usepackage{tabularx}
\usepackage{multirow}
\usepackage{graphicx}
\usepackage{xcolor}
\usepackage{colortbl}
\usepackage{tikz}
\usetikzlibrary{arrows.meta,positioning,fit}
\usepackage{etoolbox}
\usepackage{float}
\usepackage{hyperref}
\usepackage{url}
\usepackage{placeins}
\usepackage{wrapfig}
\hypersetup{hidelinks}

\newcommand{\method}{\textsc{JET}}
\newcommand{\webshop}{\textsc{WebShop}}
\newcommand{\pusht}{\textsc{PushT}}
\newcommand{\near}{\texttt{near\_success}}

\title{JET: Judge-Guided Evolution at Test Time\\for Agent Programs}
\author{
Yao Long Teng$^{1}$, Jiayi Cai$^{1}$, Bo An$^{1}$ \\
$^{1}$Nanyang Technological University
}

\iclrpreprintcopy

\begin{document}
\maketitle

\begin{abstract}
An agent’s executable program governs how it uses tools, processes observations, and responds to failures. Evolving this program at test time can help adaptation, but deciding which changes to retain is difficult when true rewards are unavailable. Execution traces provide evidence of agent behavior, yet interpreting that evidence requires a judge that remains useful as tasks and candidate programs change. We introduce Judge-Guided Evolution at Test Time (JET), which evolves an executable judge on labeled source trajectories, then freezes and transfers it to guide target-side program evolution. The judge supplies scores and diagnostic feedback without target evaluator access or model-weight updates. On unseen WebShop tasks, JET achieves approximately 13\% higher mean reward than fixed-rubric guidance when evolution begins from an unevolved program (cold start) and 4\% higher when it begins from one already optimized on source tasks (warm start), with a 36\% relative improvement in cold-start exact success. An exact-judge control on PushT, where the judge reconstructs the scoring rule from observations, shows that without judge error, program search becomes the bottleneck. Analyses identify useful reward-prediction logic in the evolved code and show that better final selection alone cannot explain the gains. These results support executable judge transfer for program adaptation under evaluator-preserving task shifts.
\end{abstract}

\section{Introduction}

Building a useful language-model agent requires decisions about the software around the model: which information to retain, which tools to invoke, and how to respond when an attempt fails \citep{yao2022react,shinn2023reflexion}.
These decisions are implemented in the agent's \emph{harness}, an executable program that organizes its interaction with a task environment. We use \emph{agent program} to cover both LLM harnesses and executable control policies.
Revising this program provides a way to change the agent's behavior while keeping the underlying model fixed.
\emph{Harness evolution} automates this process by proposing code changes, executing candidate programs, and using feedback to guide further edits \citep{lee2026meta}.
It fits the broader perspective of \emph{heuristic learning}, in which executable procedures improve through edits informed by rewards, tests, and execution records \citep{weng2026learning_beyond_gradients}. 

Such a search relies on evaluation to decide which changes to retain and what to revise next.
Task rewards provide a direct measure of candidate performance, but at test time an agent may only be able to execute programs and inspect their behavior without access to these rewards.
A judge must then infer task completion from observable behavior and provide useful feedback about remaining errors \citep{pmlr-v267-zhuge25a}.
If it systematically undervalues partial progress or misidentifies a failure, its scores and diagnostics can direct search away from useful changes. This motivates our central research question:

\noindent
\begin{tikzpicture}
\node[fill=gray!6, draw=black, line width=0.6pt, rounded corners=4pt,
      inner xsep=10pt, inner ysep=6pt, outer sep=0pt,
      text width=\dimexpr\linewidth-20pt-0.6pt\relax, align=left,
      font=\bfseries\itshape]
      {Can a judge developed on source trajectories with known rewards guide program evolution on target tasks whose evaluator outputs are withheld?};
\end{tikzpicture}
\par
Criteria for assessing behavior may remain useful even when the actions needed for success change.
Prior reward-model work provides evidence for this possibility \citep{xia2025agentrm,fu2017learning,gleave2020quantifying}.
Our focus is an executable judge whose evidence extraction, LLM assessment, and score adjustment can themselves be revised through program evolution.
The challenge is twofold: the judge must transfer across the source--target shift and remain useful as program search changes the behaviors it evaluates \citep{gao2023scaling}.
We evaluate this transfer through both reward prediction and the target-side program improvements it supports.

We introduce \method{}, a two-stage procedure that first evolves an executable judge on labeled source trajectories and selects it using a separate source validation split.
The judge is then frozen and supplies scores and diagnostics for target-side program evolution.
Underlying model weights remain fixed, and target evaluator outputs are used only for post-hoc evaluation.

\begin{figure}[t]
\centering
\includegraphics[width=\linewidth]{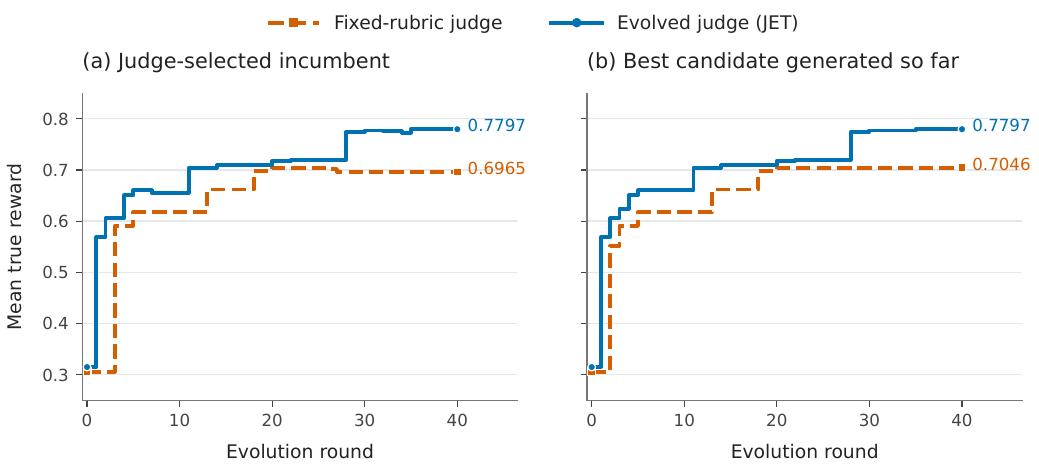}
\caption{\textbf{Judge-guided evolution on \webshop{}.} Mean true reward across five independent cold-start campaigns at each round for (a) the incumbent and (b) the best candidate generated so far within each campaign. Rewards are audited post hoc; Appendix~\ref{app:dynamics} details the aggregation.}
\label{fig:webshop-evolution-staircase}
\end{figure}

\webshop{} supplies the main test of judge transfer: a judge developed on clothing trajectories guides adaptation to non-clothing shopping tasks, and the selected programs are also evaluated on previously unseen target sessions.
Figure~\ref{fig:webshop-evolution-staircase} previews the primary cold-start searches: evolved-judge guidance reaches higher observed mean reward for both the selected program and the best candidate generated so far.
\pusht{} serves as an exact-judge control: the judge reconstructs the unchanged geometric evaluation rule from observations, which isolates program search from judge error.
Both benchmarks preserve the evaluation rule across the source--target shift; they examine adaptation when tasks or dynamics change while the scoring criterion remains fixed.

Our contributions are:
\begin{itemize}
    \item \textbf{Executable judge evolution for test-time adaptation.} We introduce \method{}, which optimizes an executable judge's evidence extraction, LLM assessment, and score adjustment on labeled source trajectories. The frozen judge then supplies scores and diagnostics for test-time agent-program evolution, without target evaluator access or updates to underlying model weights.
    \item \textbf{Evaluation under evaluator-preserving shifts.} On unseen \webshop{} tasks, JET improves mean reward over fixed-rubric guidance by approximately 13\% when starting from an unevolved program and 4\% when starting from a program optimized on source tasks. From the unevolved initialization, exact success improves by 36\% relative to fixed-rubric guidance.
    \item \textbf{Analysis of how evaluation shapes program search.} Source-holdout controls identify executable reward-prediction rules. Primary campaign results show that better final selection alone cannot explain the mean performance gap between guidance conditions. Complementary online comparisons examine executable evolution beyond prompt optimization and the roles of textual diagnostics and numerical score adjustments.
\end{itemize}

\section{Related Work}

\paragraph{Program optimization and heuristic learning.}
DSPy and Promptbreeder optimize language-model pipelines and prompts \citep{khattab2023dspy,fernando2024promptbreeder}; ADAS, STOP, and AFlow search executable agents, optimizer code, or workflows \citep{hu2024automated,zelikman2024self,zhang2025aflow}.
AlphaEvolve and GEPA further illustrate code- and prompt-based improvement from evaluation feedback \citep{novikov2025alphaevolve,agrawal2025gepa}.
The heuristic-learning perspective describes a broader process of maintaining software through code edits informed by rewards, tests, logs, and replays \citep{weng2026learning_beyond_gradients}.
We study the evaluation component of this process: whether source supervision can produce an executable judge that remains useful when the target evaluator is inaccessible.

\paragraph{Harness generation.}
Meta-Harness searches harness code using prior programs, execution traces, and task-specific scores on its search tasks \citep{lee2026meta}.
JIT-Agent trains a model to generate and repair task-conditioned harnesses \citep{zhang2026jit}.
Concurrent TTHE work studies test-time harness evolution using unlabeled execution feedback \citep{nie2026tthe}. We focus on source-supervised evolution of the evaluation procedure itself: executable judge code is optimized against known source outcomes, then frozen and transferred to guide test-time program search.

\paragraph{Learned judges and reward transfer.}
LLM judges and generative verifiers assess outputs or reasoning trajectories \citep{zheng2023judging,zhang2025generative}, while Similar learns step-wise, multidimensional signals for interactive agents \citep{miao2025boosting}.
AgentRM explicitly studies reward-model transfer and uses learned rewards for best-of-$N$ and beam search on agent tasks \citep{xia2025agentrm}.
Earlier work also learns reward models that transfer across changed initial-state distributions \citep{reddy2020learning}.
These works establish evaluation transfer as a useful direction.
Our focus is evolving the judge as executable software and using it to guide persistent program rewrites across tasks.
The judge combines evidence extraction, LLM assessment, and score adjustment, and returns textual diagnostics, while its underlying model weights remain fixed.


\section{Method: Evolving and transferring the judge}\label{sec:method}

\subsection{Problem Setting}

We consider a task space $\mathcal{T}$ and a space $\mathcal{X}$ of trajectories observable during ordinary agent execution.
An agent program is an executable program $\pi$ that maps a task $\tau\in\mathcal{T}$ to an observable trajectory $x=\pi(\tau)\in\mathcal{X}$.
An evaluator maps a task--trajectory pair to a scalar reward $R(\tau,x)\in\mathbb{R}$, which we call the true reward.
Evaluator outputs are available for labeled source trajectories used to develop the judge, but are withheld on target tasks during adaptation.
We include the deployment-visible task specification in $x$, so the judge can assess behavior against the requested goal.

A \emph{proposer} is a code-generating model that produces candidate
rewrites from the current program and available feedback.
Selecting among these candidates requires an evaluation signal.
Directly evaluated search uses $R$, whereas we study test-time search
without querying the target environment evaluator.
To guide search under this restriction, we develop a judge on labeled
source tasks $\mathcal{T}_{\mathrm{src}}$, where evaluator outputs are
available, and transfer it to reward-hidden target tasks
$\mathcal{T}_{\mathrm{adapt}}=\{\tau_i\}_{i=1}^{n}$.
Here, \emph{reward-hidden} means that evaluator outputs are withheld;
the evaluation rule need not be unknown or unreconstructible.
During target adaptation, the proposer, transferred judge, and
selection rule may access target-task instructions and
deployment-visible trajectories.
Their access excludes target evaluator outputs and derived fields,
which remain outside the adaptation workspace and are used only
for post-hoc evaluation.

Our objective is to maximize mean true reward on $\mathcal{T}_{\mathrm{adapt}}$.
Because target trajectories inform program search, performance on this set measures adaptation to the observed tasks rather than generalization to unseen tasks \citep{wang2026rethinking}.
Candidate checks enforce the permitted observation and execution interfaces.
The procedure requires executable candidates, serializable trajectories, source evaluator access, and repeatable target execution; implementation details appear in Appendix~\ref{app:implementation}.

\begin{figure}[t]
\centering
\begin{tikzpicture}[
    box/.style={draw,rounded corners,align=center,minimum height=8mm,text width=24mm,inner sep=3pt,fill=gray!7},
    arrow/.style={-{Latex[length=1.8mm]},thick},
    font=\footnotesize]
\node[anchor=west,font=\bfseries] at (-1.3,3.0) {Stage 1: Source-supervised judge evolution};
\node[box] (src) at (0,2.35) {Labeled source\\trajectories};
\node[box] (evolve) at (4.7,2.35) {Evolve judge code\\and validate};
\node[box,fill=blue!8] (frozen) at (9.4,2.35) {Freeze selected\\judge program};
\draw[arrow] (src) -- (evolve);
\draw[arrow] (evolve) -- (frozen);
\node[anchor=west,font=\bfseries] at (-1.3,1.55) {Stage 2: Reward-hidden target adaptation};
\node[box] (propose) at (0,0.65) {Propose program\\rewrite};
\node[box] (roll) at (3.13,0.65) {Execute on\\target tasks};
\node[box,fill=blue!8] (judge) at (6.26,0.65) {Judge target\\trajectories};
\node[box] (select) at (9.4,0.65) {Promote by mean\\judge score};
\draw[arrow] (propose) -- (roll);
\draw[arrow] (roll) -- (judge);
\draw[arrow] (judge) -- (select);
\draw[arrow] (frozen.south) -- ([yshift=6mm]judge.north) -- (judge.north);
\draw[arrow] (select.south) -- ++(0,-0.9) -| (propose.south);
\node[box,fill=red!5] (audit) at (3.13,-1.15) {Environment evaluator\\(audit only)};
\draw[arrow,dashed] (roll.south east) -- ++(0.1,-0.35) -- (audit.east);
\node[fill=white,font=\scriptsize] at (4.7,-0.25) {retained program + scores, trajectories, and diagnostics};
\end{tikzpicture}
\caption{Two-stage evaluator transfer. Source labels guide judge-program development. The frozen judge then supplies scores for promotion and diagnostics for subsequent program rewrites. Feedback from rejected candidates is retained. Environment-provided target evaluator outputs are isolated for auditing and have no return path to adaptation.}
\label{fig:method}
\end{figure}
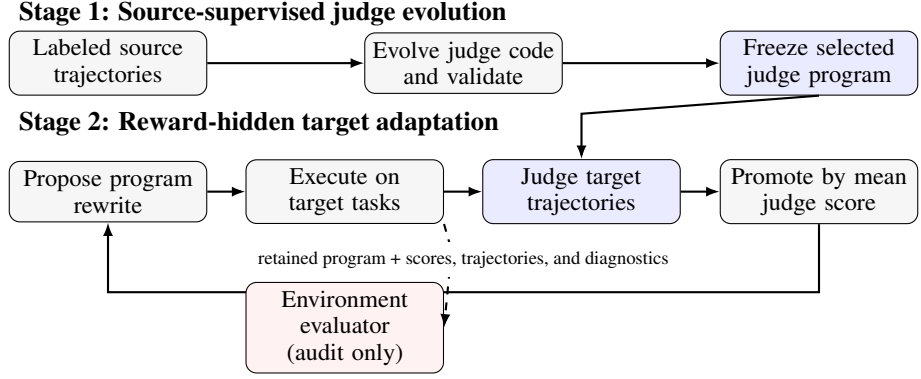

\subsection{Stage 1: Evolve an Executable Judge}

A judge $J$ maps an observable trajectory $x\in\mathcal{X}$ to a scalar estimate $\hat r=J_{\mathrm{score}}(x)$ and may additionally produce structured diagnostics $J_{\mathrm{diag}}(x)$ for the proposer.
We represent $J$ as an executable program whose prompts, trajectory preprocessing, deterministic computations, calibration logic, and output parsing may be revised during source-side evolution. The weights of any underlying inference models remain fixed.

Given labeled source trajectories $\mathcal{D}=\{(x_i,r_i)\}$, judge evolution minimizes the mean absolute error between predicted scores $J_{\mathrm{score}}(x_i)$ and source evaluator rewards $r_i$.
Source optimization supervises scalar reward prediction while diagnostic usefulness is examined through the downstream feedback interventions in Section~\ref{sec:online-evaluator-ablations}.

At each round, a proposer receives the current judge code and source-task feedback and emits a candidate judge program.
Training loss determines the next evolutionary parent and loss on a private validation split selects the final candidate after a fixed search budget.
A source holdout excluded from judge development is used for post-selection reporting.
The selected judge is then frozen: its code, parameters, and calibration logic cannot change during target adaptation.
Any underlying LLM continues to operate with the frozen judge's inference procedure.
\subsection{Stage 2: Freeze the Judge and Evolve the Agent Program}

Let $\pi_t^\star$ denote the highest-scoring eligible program observed through round $t$.
At each round, the proposer receives the code and judge score of the current best program $\pi_{t-1}^\star$ together with the deployment-visible trajectories and frozen-judge feedback from the most recently evaluated candidates.
The latter feedback remains available even when those candidates were rejected, so the current best program and the proposer's feedback state need not describe the same program.
Using this feedback, the proposer generates a set of candidate rewrites $\mathcal{C}_t$.
Each candidate is executed on all target tasks and scored by
\begin{equation}
    S_J(\pi)=\frac{1}{n}\sum_{i=1}^{n}J_{\mathrm{score}}\!\left(\pi(\tau_i)\right).
    \label{eq:judge-score}
\end{equation}
Let $\mathcal{C}_t^{\mathrm{elig}}\subseteq\mathcal{C}_t$ contain candidates satisfying all interface and, when enabled, source constraints; when this set is nonempty, let $\bar\pi_t$ be its highest-scoring member.
The current best program is updated only when $\bar\pi_t$ has a higher judge score:
\begin{equation}
    \pi_t^\star=
    \begin{cases}
        \bar\pi_t, & \mathcal{C}_t^{\mathrm{elig}}\neq\varnothing
        \text{ and }S_J(\bar\pi_t)>S_J(\pi_{t-1}^\star),\\
        \pi_{t-1}^\star, & \text{otherwise}.
    \end{cases}
\end{equation}
Selection in the proposed reward-hidden conditions uses no target evaluator output.
The initial program may be a cold seed program or a warm program previously optimized using labeled source tasks.
The transferred judge supplies scores and diagnostics: its scores determine promotion and enter proposer feedback, while its diagnostics and the trajectories of both accepted and rejected candidates also inform subsequent rewrites.

\subsection{From Reward Prediction to Search Guidance}

The judge learns to assess realized behavior, while target-side search discovers how to produce it.
Its average prediction error is only one measure of usefulness \citep{frick2025evaluate,lou2025rethinking}: promotion depends on the ordering of candidate mean scores in Equation~\ref{eq:judge-score}.
For two candidates with different true mean rewards, a useful judge assigns the higher score to the better candidate.
Agreement near promotion decisions is therefore relevant even when overall prediction error is small. The core procedure requires no source-evaluator queries during target adaptation.
An optional source constraint can restrict eligible candidates to those meeting an evaluator-measured floor on a fixed source check set (Appendix~\ref{app:source-floor}).

\section{Experimental Setup}\label{sec:setup}

\subsection{Tasks and Transfer Settings}

\webshop{} is used to test interactive shopping under a clothing-to-non-clothing shift \citep{yao2022webshop}.
Clothing source tasks are divided into training, validation, and holdout sets for judge development, selection, and evaluation.
Adaptation uses non-clothing tasks and all splits are session-disjoint. Additionally, held-out tasks never guide adaptation.

\pusht{} tests continuous control by pushing a T-shaped block into a goal \citep{chi2025diffusion}.
Source tasks begin near the goal without damping; target tasks have larger initial displacement and nonzero damping.
The geometric evaluation rule is unchanged and exactly reconstructible from visible poses, making this an exact-judge control.
Dataset construction and execution settings are collected in Appendix~\ref{app:implementation}.

\subsection{Comparison Conditions}

We compare no target evolution, proposer self-selection, and guidance from the fixed-rubric or evolved judge.
The fixed-rubric judge uses a predefined task-specific rubric and scoring procedure, without source-label optimization.
For \webshop{} proposer self-selection, the proposer assigns each candidate a scalar self-score after reading its reward-scrubbed trajectories, with the candidate with the highest self-score being selected.
The evolved judge is developed on source trajectories and selected by source validation MAE, and both judges remain frozen during target adaptation.
In the primary \webshop{} experiments, the seed judge that initializes evolution is distinct from the fixed-rubric judge.
A \emph{warm start} uses a source-optimized program; a \emph{cold start} uses the unevolved seed.
Each primary condition is run as five independent campaigns; a campaign is one complete 40-round evolution run from a given initialization.
We report campaign means and standard deviations.
Within each comparison and initialization, adaptive conditions use the same task sets and test-time proposal budgets.
Under full judge guidance, mean scalar estimates determine promotion, while scores and textual diagnostics inform rewrites. Source-constrained conditions, labeled \emph{source constraint} in Table~\ref{tab:main-results}, additionally query labeled source tasks to enforce a fixed performance floor (Appendix~\ref{app:source-floor}).

\subsection{Evaluation Metrics}

Primary program metrics are mean true reward in \webshop{} and mean final goal coverage in \pusht{}, the fraction of goal area overlapped by the block.
Judge fidelity is measured by MAE and correlation with true rewards.
Environment-provided target evaluator outputs are withheld during adaptation. In \pusht{}, the frozen judge nevertheless reconstructs the final-coverage metric exactly from visible poses.
For each run, \emph{oracle-best} is the highest true performance among its generated candidates; \emph{selection regret} is the gap between that performance and the selected candidate's audited performance.
These diagnostics are computed within each campaign and then averaged; source-constrained accounting is given in Appendix~\ref{app:source-floor}.

\section{Results}\label{sec:results}
\FloatBarrier
\begin{table}[t]
\centering
\caption{Test-time evolution across five independent campaigns. Panel A: \webshop{} reward; Panel B: \pusht{} coverage (exact-judge control). Selected values are mean $\pm$ SD; other columns are means. Bold marks the best selected mean and gain per benchmark and initialization; grey denotes \method{}. Environment evaluator outputs are withheld online; the \pusht{} judge reconstructs coverage exactly.}
\label{tab:main-results}
\setlength{\tabcolsep}{6pt}
\begin{tabularx}{\linewidth}{lXrrrr}
\toprule
Start & Selection signal & Initial & Selected & $\Delta$ & Regret $\downarrow$ \\
\midrule
\multicolumn{6}{l}{\textbf{Panel A: \webshop{}}} \\
\midrule
Warm & None & 0.6256 & $0.6256\pm0.0175$ & -- & -- \\
Warm & Fixed-rubric judge & 0.6256 & $0.6589\pm0.0299$ & +0.0333 & 0.0198 \\
\rowcolor{gray!10}
Warm & Evolved judge & 0.6256 & $0.6951\pm0.0706$ & +0.0694 & 0.0000 \\
\rowcolor{gray!10}
Warm & Evolved + source constraint & 0.6250 & $\mathbf{0.7034}\pm0.0122$ & \textbf{+0.0784} & 0.0111 \\
\midrule
Cold & Self-selection & 0.3043 & $0.6137\pm0.0699$ & +0.3094 & 0.0121 \\
Cold & Fixed-rubric judge & 0.3059 & $0.6965\pm0.0268$ & +0.3906 & 0.0081 \\
\rowcolor{gray!10}
Cold & Evolved judge & 0.3150 & $\mathbf{0.7797}\pm0.0702$ & \textbf{+0.4647} & 0.0000 \\
\rowcolor{gray!10}
Cold & Evolved + source constraint & 0.3158 & $0.7100\pm0.0109$ & +0.3942 & 0.0000 \\
\midrule
\multicolumn{6}{l}{\textbf{Panel B: \pusht{}}} \\
\midrule
Warm & None & 0.3709 & $0.3709\pm0.0008$ & -- & -- \\
Warm & Fixed-rubric judge & 0.3709 & $0.6033\pm0.0087$ & +0.2324 & 0.0475 \\
\rowcolor{gray!10}
Warm & Evolved judge & 0.3709 & $\mathbf{0.6683}\pm0.0459$ & \textbf{+0.2974} & 0.0000 \\
\midrule
Cold & Self-selection & 0.0000 & $0.6306\pm0.1352$ & +0.6306 & 0.0628 \\
Cold & Fixed-rubric judge & 0.0000 & $0.6199\pm0.0624$ & +0.6199 & 0.0306 \\
\rowcolor{gray!10}
Cold & Evolved judge  & 0.0000 & $\mathbf{0.7679}\pm0.0157$ & \textbf{+0.7679} & 0.0000 \\
\bottomrule
\end{tabularx}
\end{table}

\subsection{WebShop Adaptation and Held-Out Transfer}

Panel A of Table~\ref{tab:main-results} reports higher observed mean rewards under evolved-judge guidance (JET) from both initializations, with a larger mean difference from the cold start.

\textbf{Warm start.}
Without test-time evolution, the warm program obtains 0.6256 mean reward.
Evolution with the fixed-rubric judge raises this to 0.6589, whereas evolution with the evolved judge reaches 0.6951.
The observed mean reward gain is approximately twice that under fixed-rubric guidance, with a final mean difference of 0.0362.
Warm-start exact success on the adaptation set is slightly lower with the evolved judge (30.6\% versus 31.2\%), despite its higher mean reward (Appendix~\ref{app:dynamics}).

\textbf{Cold start.}
The observed mean separation is larger when evolution begins from the unevolved program.
Proposer self-selection reaches 0.6137 mean reward, guidance from the fixed-rubric judge reaches 0.6965, and guidance from the evolved judge reaches 0.7797.
The evolved-judge condition has a higher observed mean reward by 0.0832 over fixed-rubric guidance and by 0.1660 over self-selection.
The optional source constraint has a slightly higher warm-start target mean and a lower cold-start target mean, while maintaining the required source performance (Appendix~\ref{app:source-floor}).

Under both judges, cold-start evolution reaches higher final mean rewards than warm-start evolution despite beginning from weaker programs. Source optimization biases subsequent rewrites toward source-specific behaviors, making adaptation to the target distribution more difficult. Starting from an unevolved program may therefore give search greater flexibility to develop target-specific behavior.

\paragraph{Held-out target tasks.}

To test whether the improvements extend beyond the tasks observed during evolution, we freeze each selected program and evaluate it on a disjoint set of previously unseen tasks sampled from the same target distribution.
These tasks are not used for evolution, candidate selection, or model development. Panel A of Table~\ref{tab:webshop-transfer-analysis} shows that the adaptation-set ordering is preserved on the held-out set.
From the cold start, evolution with the evolved judge reaches 0.7915 reward, compared with 0.7008 under the fixed-rubric judge and 0.6259 under proposer self-selection.
From the warm start, evolution with the evolved judge reaches 0.6951, compared with 0.6700 under the fixed-rubric judge and 0.6387 without test-time evolution.
Cold-start evolution with the evolved judge also yields the highest mean held-out exact-success rate (50.6\%).

\begin{table}[t]
\centering
\caption{\webshop{} held-out performance and offline controls. (A) Reward mean $\pm$ SD and mean success; \emph{no external judge} denotes no evolution from warm and self-selection from cold. (B) Source-holdout MAE and five-campaign mean target MAE and selection regret. Target columns identify the guidance used to generate candidates.}
\label{tab:webshop-transfer-analysis}
\setlength{\tabcolsep}{5pt}
\begin{tabularx}{\linewidth}{Xrrrr}
\toprule
\multicolumn{5}{@{}l}{\textbf{Panel A: Primary held-out program performance}} \\
\addlinespace[2pt]
& \multicolumn{2}{c}{Warm start} & \multicolumn{2}{c}{Cold start} \\
\cmidrule(lr){2-3}\cmidrule(l){4-5}
Adaptation guidance & Reward $\uparrow$ & Success $\uparrow$ & Reward $\uparrow$ & Success $\uparrow$ \\
\midrule
No external judge & $0.6387\pm0.0021$ & 27.6\% & $0.6259\pm0.0354$ & 24.8\% \\
Fixed-rubric judge & $0.6700\pm0.0374$ & 32.6\% & $0.7008\pm0.0499$ & 37.2\% \\
\rowcolor{gray!10}
Evolved judge & $\mathbf{0.6951}\pm0.0052$ & \textbf{34.2\%} & $\mathbf{0.7915}\pm0.0322$ & \textbf{50.6\%} \\
\bottomrule
\end{tabularx}

\vspace{6pt}
\begin{tabular*}{\linewidth}{@{}l@{\extracolsep{\fill}}rrrrr@{}}
\toprule
\multicolumn{6}{@{}l}{\textbf{Panel B: Offline judge controls}} \\
\addlinespace[2pt]
& \multicolumn{3}{c}{Prediction error (MAE $\downarrow$)}
& \multicolumn{2}{c}{Selection regret $\downarrow$} \\
\cmidrule(lr){2-4}\cmidrule(l){5-6}
Scoring judge & Source & Fixed-rubric & Evolved & Fixed-rubric & Evolved \\
\midrule
Fixed-rubric judge & 0.1576 & 0.2252 & 0.2239 & 0.0081 & 0.0040 \\
Calibrated fixed-rubric judge & 0.1488 & 0.2227 & 0.2250 & 0.0081 & 0.0040 \\
Evolved judge & 0.1168 & 0.1682 & 0.1843 & 0.0000 & 0.0000 \\
Deterministic evolved judge & 0.1153 & 0.1533 & 0.1695 & 0.0000 & 0.0000 \\
\bottomrule
\end{tabular*}
\end{table}

\subsection{What the Judge Learns}\label{sec:judge-analysis}

\begin{wrapfigure}{R}{0.50\linewidth}
\centering
\includegraphics[width=\linewidth]{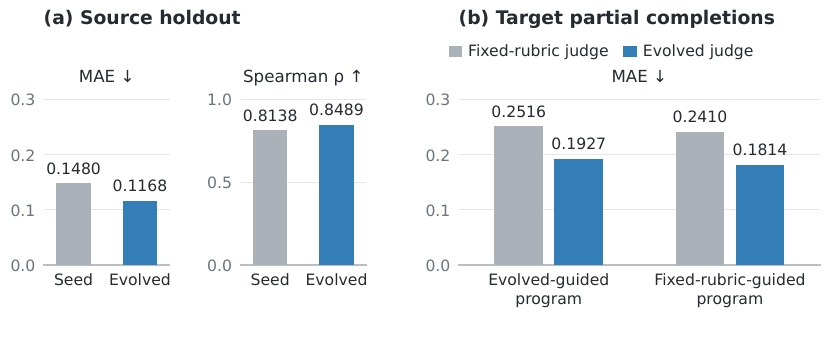}
\caption{Judge fidelity. (a) Source-holdout MAE and Spearman $\rho$: seed versus evolved. (b) Partial-reward MAE: fixed-rubric versus evolved on identical trajectories from each auxiliary program. Lower MAE and higher $\rho$ are better.}
\label{fig:source-aux-fidelity}
\end{wrapfigure}
Figure~\ref{fig:source-aux-fidelity}(a) shows that source evolution reduces the seed judge's holdout MAE from 0.1480 to 0.1168 (21.1\%) and raises Spearman correlation by 0.0351. These gains indicate better reward estimates and more accurate trajectory ranking on source data excluded from development.
Figure~\ref{fig:source-aux-fidelity}(b) compares both judges on identical trajectories from two auxiliary programs, one adapted under each judge. On partially completed tasks, the evolved judge reduces reward-prediction MAE by 23.4\% for the evolved-guided program and 24.7\% for the fixed-rubric-guided program. The accuracy advantage therefore also appears on trajectories produced under fixed-rubric guidance. Appendix~\ref{app:error} shows less underestimation of partial progress, although the fixed-rubric judge remains more accurate on complete failures.

To examine what judge evolution contributes, we compare the fixed-rubric judge with a source-calibrated version, the evolved judge, and a deterministic version that runs evidence extraction and score adjustment without LLM assessment.

\paragraph{Can score calibration match evolved-judge accuracy?}
Fitting an isotonic mapping \citep{zadrozny2002transforming} on labeled source trajectories reduces source-holdout MAE from 0.1576 to 0.1488, compared with 0.1168 for the evolved judge.
The evolved judge also has lower MAE under both target guidance conditions (Table~\ref{tab:webshop-transfer-analysis}, Panel B).
Calibration and scoring-control details appear in Appendix~\ref{app:offline-evaluator-controls}.

\paragraph{Can evolved code predict rewards without fresh LLM assessment?}
The deterministic evolved judge achieves source-holdout MAE of 0.1153, compared with 0.1168 for the full judge, and lower MAE on both cached target pools (Table 2, Panel B). Comparison with the seed judge’s code shows that this independent numerical pathway emerged during source evolution. These results indicate that the evolved code captures useful reward-prediction rules, allowing it to maintain prediction accuracy on these trajectory sets without a fresh LLM assessment. The full judge additionally generates textual diagnostics for program rewrites, whose contribution is examined separately in Section 5.4.

\begin{figure}[!ht]
\centering
\setlength{\fboxsep}{5pt}
\colorbox{gray!8}{\parbox{\dimexpr\linewidth-2\fboxsep\relax}{\small
\textbf{Request:} ivory/blue rug, $2$~ft $3$~in $\times$ $6$~ft, below \$130.\\
\textbf{Purchase:} matching rug and color, $2$~ft $\times$ $8$~ft, \$100.\\
\textbf{True reward:} 0.8333; wrong size prevents exact success.}}
\par\vspace{5pt}
\begin{minipage}[t]{0.487\linewidth}
\colorbox{orange!8}{\parbox{\dimexpr\linewidth-2\fboxsep\relax}{\small
\textbf{Fixed-rubric judge}\\
Predicted reward: \textbf{0.3200}\\
Finds the size mismatch but also raises a spurious price concern.}}
\end{minipage}\hfill
\begin{minipage}[t]{0.487\linewidth}
\colorbox{blue!7}{\parbox{\dimexpr\linewidth-2\fboxsep\relax}{\small
\textbf{Evolved judge}\\
Predicted reward: \textbf{0.8329}\\
Credits the product, selected color, and price; localizes the size error.}}
\end{minipage}
\caption{Selected partial-progress example. Both judges score the same reward-scrubbed trajectory; true reward is shown only post hoc. The evolved judge credits satisfied requirements and identifies the remaining error.}
\label{fig:motivation}
\end{figure}

\paragraph{How do scoring and diagnosis differ on the same trajectory?}
Figure~\ref{fig:motivation} compares the judges on a purchase that satisfies the requested product, color, and price but selects the wrong size.
The fixed-rubric judge identifies the size mismatch, yet substantially underestimates partial completion and raises an unsupported price concern.
The evolved judge credits the satisfied requirements and identifies size as the remaining mismatch. Its more accurate partial-credit estimate is consistent with the lower aggregate partial-reward error in Figure~\ref{fig:source-aux-fidelity}(b).

These differences matter because scores and diagnostics serve different roles in program evolution.
Scores determine which candidates are retained and also enter proposer feedback; diagnostics provide additional guidance for subsequent revisions \citep{yuksekgonul2025optimizing,madaan2023self}.
Undercrediting completed requirements can undervalue a candidate while an unsupported criticism can direct revisions toward an already satisfied requirement.

\subsection{Selection Quality and Candidate Generation}

\paragraph{Does improved selection extend to intermediate rounds?}
At each round within a campaign, each judge selects from the saved candidates generated so far, holding their trajectories fixed.
Under fixed-rubric candidate generation, the evolved judge has zero selection regret at every round, whereas the fixed-rubric judge incurs regret at a mean of 17 of 40 rounds per campaign.
On candidates generated under evolved-judge guidance, the fixed-rubric and evolved judges incur regret at an average of 18 and 8 rounds per campaign, respectively, and mean campaign maximum regret falls from 0.2538 to 0.0185.
The selection advantage therefore extends beyond the final choice under both guidance conditions.

\paragraph{Can better final selection alone explain the performance gain?}
For each campaign, we hold the saved programs fixed and compare which program each judge selects. The evolved judge achieves zero mean final regret on candidates generated under both fixed-rubric guidance and evolved-judge guidance, while calibration leaves the fixed-rubric judge’s mean selected reward unchanged (Table 2, Panel B). Perfect final selection among programs generated under fixed-rubric guidance would raise mean true adaptation reward from 0.6965 to approximately 0.7046, still below 0.7797 under evolved guidance. The mean reward advantage of evolved-judge-guided search would therefore shrink from 0.0832 to 0.0751, closing only about 9.7\% of the original difference. Most of this advantage thus reflects differences in the programs generated during search, which better final selection alone cannot recover.

\subsection{Judge Baselines and Component Ablations}
\label{sec:online-evaluator-ablations}

\begin{wrapfigure}{R}{0.50\linewidth}
\centering
\includegraphics[width=\linewidth]{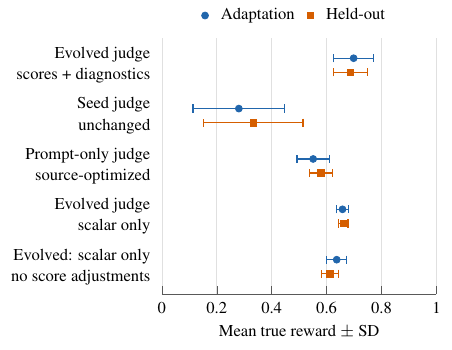}
\caption{Judge and feedback comparisons (mean $\pm$ SD). See Appendix~\ref{app:online-ablation-results}.}
\label{fig:online-evaluator-ablations}
\end{wrapfigure}
Figure~\ref{fig:online-evaluator-ablations} compares judge development and feedback components through fresh program evolution in a ablation series. Held-out reward is measured on separate target tasks using frozen selected programs.

\textbf{Does evolving executable judge code offer benefits beyond optimizing its prompt?}
The source-optimized prompt-only judge holds its executable wrapper fixed. Both source-optimized judges achieve higher observed mean rewards than the unchanged seed, but executable evolution exceeds prompt optimization by 0.1473 in adaptation reward and 0.1074 in held-out reward. Prompt refinement leaves the judge’s evidence extraction and score adjustment code unchanged, whereas executable evolution can revise these components alongside its instructions. The results support optimizing the evaluation procedure as a whole, with the held-out advantage indicating that the resulting gains extend beyond the tasks used for adaptation.

\textbf{What changes when textual diagnostics are included?}
With the evolved judge's scoring procedure held fixed, mean adaptation reward is 0.6984 with diagnostics and 0.6579 with scalar feedback only while mean held-out reward is 0.6869 and 0.6608 respectively.
These mean differences provide evidence for a benefit from diagnostics. Diagnostics identify requirements and suggest edits beyond the scalar completion estimate.

\textbf{Do numerical score adjustments contribute to program performance?}
With diagnostics withheld, we bypass we bypass score adjustment while retaining evidence extraction and LLM assessment. Retaining these adjustments yields higher observed mean adaptation and held-out rewards by 0.0211 and 0.0484, respectively. These results suggest that the evolved score adjustment makes scalar feedback more useful for program adaptation beyond the underlying LLM assessment. Because scores determine candidate promotion and also inform subsequent rewrites, the adjustments can influence both which programs are retained and how later candidates are generated.

\subsection{PushT: An exact-judge control}

We use \pusht as an exact-judge control that isolates program search from judge error.
The evolved judge recovers the geometric coverage rule in the first source-evolution round and computes block--goal overlap from observable terminal poses (Appendix~\ref{app:pusht}).
Its scalar selection signal therefore equals the the true final coverage metric, although environment-provided reward fields remain withheld.

All adaptation conditions achieve higher observed mean coverage than their starting policies (Table 1, Panel B). From cold, evolved-judge guidance reaches 0.7679 coverage, compared with 0.6199 under fixed-rubric guidance and 0.6306 under proposer self-selection. From warm, it reaches 0.6683 versus 0.6033 under fixed-rubric guidance, an advantage of 0.0650. Because the evolved judge is exact, the gap to fixed-rubric guidance reflects what an approximate judge costs the search. Zero selection regret under evolved-judge guidance means that each campaign selects a policy with the highest true coverage among its generated candidates. Yet mean coverage remains below one even with exact scoring. Further improvement therefore requires generating better policies, rather than making a different final choice among the same candidates.
\FloatBarrier
\section{Conclusion and Limitations}
\label{sec:conclusion}
\label{sec:limitations}

\method{} evolves an executable judge on labeled source trajectories and transfers its frozen scores and diagnostics to test-time program search without target evaluator access.
Adapted programs attain higher observed mean rewards, including on unseen target tasks, but limited campaign replication leaves smaller differences, including diagnostic-feedback gains, uncertain.
Both benchmarks preserve the evaluation rule; transfer to changed reward criteria or outcomes requiring unobserved information remains untested.
Repeated program generation, execution, and evaluation incur inference cost and latency. Broader replication beyond the five primary campaigns and cost analysis remain future work.

\subsection*{AI use statement}

In this work, we used generative AI tools for feedback on our experimental methodology
and on the interpretation of results. We have not used generative AI tools for developing
the method or its hypotheses, generating or cleaning data, and
mathematical claims, proofs, translation, and qualitative data analysis are not applicable
to this work. Additionally, we used generative AI tools for readability editing, drafting
and restructuring text, and formatting references and LaTeX.
We take responsibility for the final content of this work,
including text, claims or artifacts produced with the aid of generative AI.

\subsection*{Reproducibility statement}
Sections~\ref{sec:setup} and~\ref{sec:results} describe the comparison protocol, metrics, and results.
Appendix~\ref{app:implementation} collects campaign replication, data splits, search budgets, and execution settings.
The remaining appendices provide source-constraint details, judge-error analyses, primary campaign diagnostics, qualitative examples, offline control procedures, and ablation results. Additionally, we will release the code base and artifacts for this research work. 

\bibliography{iclr2025_conference}
\bibliographystyle{iclr2027_conference}

\appendix
\section{Experimental Settings}
\label{app:implementation}

\subsection{Independent Campaigns and Search Settings}
\label{app:campaign-settings}

Each primary test-time adaptation condition is evaluated using five independent campaign seeds.
A campaign is a complete program-evolution run from a specified initialization; evolution rounds and candidate proposals are steps within that run.
Within each campaign, the judge remains frozen and environment-provided evaluator outputs are used only for post-hoc measurement.
Dataset-shuffling seeds are distinct from campaign replication.

Table~\ref{tab:primary-settings} collects the primary test-time search settings.
For each selected program, reward and success are first averaged over its evaluation tasks.
Initial reward, improvement, oracle-best reward, and selection regret are computed within each campaign before averaging across campaigns.
Tables~\ref{tab:main-results} and~\ref{tab:webshop-transfer-analysis} (Panel A) report campaign means, with SDs for selected-program rewards and coverage where available. Panel B reports source-holdout prediction error and campaign-averaged target prediction error and selection regret. The code proposer used is GPT-5.5 through Codex, with medium reasoning effort while both solver’s and judge's LLM calls use GPT-OSS-120B, served locally. Figure~\ref{fig:online-evaluator-ablations} and Table~\ref{tab:online-evaluator-ablations} report campaign means and standard deviations for the judge and feedback comparisons.
Offline scoring and selection procedures are specified in Appendix~\ref{app:offline-evaluator-controls}.

\begin{table}[htbp]
\centering
\caption{Primary test-time search settings. These settings are shared by adaptive conditions within each benchmark and initialization. Baselines without test-time evolution perform no search.}
\label{tab:primary-settings}
\begin{tabular*}{\linewidth}{@{}l@{\extracolsep{\fill}}cc@{}}
\toprule
Setting & \webshop{} & \pusht{} \\
\midrule
Independent campaign seeds per condition & 5 & 5 \\
Evolution iterations per campaign & 40 & 40 \\
Maximum environment steps per trajectory & 100 & 100 \\
Initializations & Warm and cold & Warm and cold \\
\bottomrule
\end{tabular*}
\end{table}

\subsection{WebShop Data and Execution Details}

This subsection describes the primary \webshop{} experiments.

Experiments use the full catalog of 1,181,436 products and an 11,674,685-goal synthetic attribute-generated universe, with upstream task-shuffle seed 233.
The adaptation set contains 500 non-clothing tasks: 300 garden, 81 beauty, 71 electronics, and 48 grocery.
The held-out target set has the same domain composition and disjoint session IDs; it is never exposed to the proposer, judge, or selection rule.

Source data comprise 500 clothing training tasks, 300 validation tasks, and 300 holdout tasks.
The evolved judge uses a fixed trajectory dataset generated by the seed program on these tasks; holdout trajectories are used only for reporting.
Judge development uses 20 proposer rounds, training-MAE candidate selection, and private validation-MAE final selection.
Warm programs are evolved on the 500 source training tasks, and the optional source constraint queries a separate fixed set of 50 source tasks.

\paragraph{Observation scrubbing.}
Trajectories supplied to the proposer and judge retain the task instruction, ordinary product observations, actions, and available interaction targets.
Environment-provided final and per-step rewards, coverage, success indicators, reward breakdowns, and evaluator-only goal annotations are excluded.
Scrubbing removes evaluator-derived information from both structured trajectory fields and rendered observation text, including the post-purchase final observation.
The retained task instruction and ordinary product information remain available for judging completion; target evaluator outputs and evaluator-only annotations are not exposed.

The judge emits a continuous \near{} score, a binary \texttt{passed} field, and textual failure, reason, and repair fields.
Mean \near{} is used as $J_{\mathrm{score}}$ for candidate promotion; the binary field is reported but does not determine promotion, while diagnostics remain visible to the proposer.
The selected judge combines an LLM call with deterministic calibration logic.

The primary proposal protocol permits use of information and vocabulary visible in the target interaction, but prohibits explicit task identifiers, hidden evaluator access, and external campaign state.
The resulting program may specialize to the observed deployment environment.
No target evaluator output is returned to the proposer, judge, or promotion rule in any reported test-time campaign.

\subsection{PushT Task and Protocol Details}
\label{app:pusht}

The agent program is an executable Python control policy that maps observed state to a two-dimensional target position.
The proposer and judge observe block and goal poses; stored coverage, success, and reward fields are withheld.
Final coverage is the intersection area of the final block and goal shapes divided by the goal area.
The fixed-rubric judge estimates completion with an LLM rubric, while the evolved judge reconstructs overlap from visible geometry.
Let $B_T\subset\mathbb{R}^2$ denote the region occupied by the block
at its terminal pose, and let $G\subset\mathbb{R}^2$ denote the goal region.
Final coverage is
\begin{equation}
C_T =
\frac{\operatorname{Area}(B_T \cap G)}
     {\operatorname{Area}(G)}
\in [0,1].
\label{eq:pusht-coverage}
\end{equation}
Source tasks start near the goal without damping; target tasks use larger initial displacement and damping 0.2.
The warm policy is developed on labeled source tasks.
The frozen warm policy obtains 0.3709 coverage on the target set, while the cold policy obtains zero.
Target adaptation uses the settings in Table~\ref{tab:primary-settings} and retains the highest-scoring eligible policy.

The fixed-rubric judge has training and validation coverage MAE of 0.0624 and 0.0742.
The evolved judge computes the final polygon intersection fraction and obtains 0.0000 MAE on both splits.
We treat this condition as an exact-judge control: policy behavior must adapt to changed dynamics, while the reconstructed coverage rule remains unchanged.

\section{Optional Source-Capability Preservation}
\label{app:source-floor}

\subsection{Constraint Definition}

The core method requires no source-evaluator queries during target adaptation.
When retaining a verified level of source capability is desirable, the framework can optionally impose a labeled source constraint without using target evaluator outputs.
Let $G\subset\mathcal{T}_{\mathrm{src}}$ be a source check set, $Q_G(\pi)$ its source-evaluator score, $b$ a fixed reference floor, and $\epsilon\ge0$ a tolerance.
In the reported experiments, $b$ is the initial program's score on the source check set.
A candidate $\pi$ is eligible only if
\begin{equation}
    Q_G(\pi)\ge b-\epsilon.
\end{equation}
The constraint anchors otherwise reward-hidden target adaptation to an evaluator-measured source capability.
It guarantees only aggregate performance on the fixed source check set, not target performance or per-task source non-regression.
Its task set, metric, reference policy, and tolerance are environment-specific experimental choices rather than required components of evaluator transfer.
For source-constrained campaigns, reported selection regret can include the cost of excluding source-ineligible candidates; it should not be interpreted solely as judging error.

\section{Detailed Error Decomposition}
\label{app:error}

Judge-quality analyses report MAE and Spearman correlation.
For completeness, the seed judge and evolved judge obtain training MAE of 0.1685 and 0.1038, respectively, and validation MAE of 0.1637 and 0.0982.
The main paper reports prediction error on the source holdout excluded from judge development.

Table~\ref{tab:source-aux-fidelity} summarizes source-holdout fidelity and partial-reward accuracy.

\begin{table}[htbp]
\centering
\caption{Judge fidelity. Source rows compare seed and evolved judges; target rows compare fixed-rubric and evolved judges on auxiliary partial-reward trajectories.}
\label{tab:source-aux-fidelity}
\setlength{\tabcolsep}{4pt}
\begin{tabular*}{\linewidth}{@{}l@{\extracolsep{\fill}}rr@{}}
\toprule
Metric / trajectory source & Reference & Evolved \\
\midrule
\multicolumn{3}{@{}l}{\textbf{Source holdout: seed reference}} \\
MAE $\downarrow$ & 0.1480 & \textbf{0.1168} \\
Spearman $\rho$ $\uparrow$ & 0.8138 & \textbf{0.8489} \\
\midrule
\multicolumn{3}{@{}l}{\textbf{Target partial-reward MAE $\downarrow$}} \\
\multicolumn{3}{@{}l}{Fixed-rubric reference} \\
Evolved-guided program & 0.2516 & \textbf{0.1927} \\
Fixed-rubric-guided program & 0.2410 & \textbf{0.1814} \\
\bottomrule
\end{tabular*}
\end{table}

The following matched cross-scoring analysis uses saved trajectories from auxiliary \webshop{} campaigns rather than the primary program matrix.
Within each row, both frozen judges score the identical 500 reward-scrubbed trajectories, so the comparison isolates prediction on a fixed trajectory set.

\begin{table}[htbp]
\centering
\caption{Auxiliary matched target cross-scoring. Panel A reports aggregate prediction and ranking accuracy; Panel B decomposes error by ground-truth reward regime. Bias is mean prediction minus true reward. Lower MAE and higher Spearman $\rho$ are better.}
\label{tab:matched-cross-scoring}
\setlength{\tabcolsep}{4pt}
\begin{tabular}{lrrrrr}
\toprule
\multicolumn{6}{l}{\textbf{Panel A: Aggregate accuracy and ranking}} \\
\midrule
Trajectory set & True mean & \shortstack{Fixed-rubric\\MAE} & \shortstack{Evolved\\MAE} & \shortstack{Fixed-rubric\\$\rho$} & \shortstack{Evolved\\$\rho$} \\
\midrule
\shortstack[l]{Program guided by\\evolved judge} & 0.7373 & 0.2824 & \textbf{0.1811} & 0.5069 & \textbf{0.6613} \\
\shortstack[l]{Program guided by\\fixed-rubric judge} & 0.7025 & 0.2377 & \textbf{0.1725} & 0.6055 & \textbf{0.7029} \\
\bottomrule
\end{tabular}
\vspace{1.5mm}

\begin{tabular}{llrrrrr}
\toprule
\multicolumn{7}{l}{\textbf{Panel B: Error decomposition by reward regime}} \\
\midrule
Trajectory set & Regime & $N$ & \shortstack{Fixed-rubric\\MAE} & \shortstack{Fixed-rubric\\bias} & \shortstack{Evolved\\MAE} & \shortstack{Evolved\\bias} \\
\midrule
\shortstack[l]{Program guided by\\evolved judge} & Zero & 15 & \textbf{0.1500} & +0.1500 & 0.3954 & +0.3954 \\
 & Partial & 287 & 0.2516 & -0.1741 & \textbf{0.1927} & -0.0042 \\
 & Exact & 198 & 0.3370 & -0.3370 & \textbf{0.1480} & -0.1480 \\
\midrule
\shortstack[l]{Program guided by\\fixed-rubric judge} & Zero & 22 & \textbf{0.1523} & +0.1523 & 0.3006 & +0.3006 \\
 & Partial & 297 & 0.2410 & -0.1521 & \textbf{0.1814} & +0.0139 \\
 & Exact & 181 & 0.2425 & -0.2425 & \textbf{0.1423} & -0.1423 \\
\bottomrule
\end{tabular}
\end{table}

Most target outcomes are partial or exact.
The evolved judge's aggregate advantage comes from these strata, while the fixed-rubric judge better recognizes the small number of true-zero cases.
This trade-off is compatible with useful dense optimization and occasional severe false optimism.

\section{Primary Campaign Diagnostics}
\label{app:dynamics}

\paragraph{Search trajectories.}
Figure~\ref{fig:webshop-evolution-staircase} compares the primary cold-start searches guided by the fixed-rubric and evolved judges.
Each campaign uses 40 valid evolution rounds, and 500 adaptation tasks, without a source constraint.
Within each campaign, candidate rewards are first averaged over the adaptation tasks.
Panel~(a) then averages the retained programs' rewards across campaigns at each round.
Panel~(b) first identifies the highest mean reward attained within each campaign's generated candidates through that round, including rejected candidates, and then averages these values across campaigns.
Hidden rewards are used only for post-hoc analysis and never for online selection.
Round~0 is evaluated separately in each campaign.
Each curve averages independent target-adaptation campaigns;
Held-out target performance is reported separately in Table~\ref{tab:webshop-transfer-analysis}, Panel~A.

Table~\ref{tab:primary-search-diagnostics} reports additional diagnostics from the primary adaptation campaigns.
Oracle-best and regret are computed within each campaign's generated candidates before averaging across campaigns.

\begin{table}[htbp]
\centering
\caption{Search diagnostics averaged across the \webshop{} campaigns. Selection score is the judge's predicted score under judge guidance and the proposer's self-score under self-selection. Oracle-best, regret, and exact success use hidden reward only post hoc. Success is measured on the adaptation tasks.}
\label{tab:primary-search-diagnostics}
\begin{tabular}{lrrrr}
\toprule
Condition & Selection score & Oracle-best & Regret & Success \\
\midrule
Warm + fixed-rubric judge & 0.5059 & 0.6788 & 0.0198 & 31.2\% \\
Warm + evolved judge & 0.6487 & 0.6951 & 0.0000 & 30.6\% \\
Warm + evolved judge + source constraint & 0.6550 & 0.7145 & 0.0111 & 35.2\% \\
Cold + self-selection & 0.7800 & 0.6258 & 0.0121 & 23.0\% \\
Cold + fixed-rubric judge & 0.5701 & 0.7046 & 0.0081 & 35.6\% \\
Cold + evolved judge & 0.7489 & 0.7797 & 0.0000 & 49.2\% \\
Cold + evolved judge + source constraint & 0.6966 & 0.7100 & 0.0000 & 35.8\% \\
\bottomrule
\end{tabular}
\end{table}

\FloatBarrier

\section{Offline Judge Control Procedures}
\label{app:offline-evaluator-controls}

\paragraph{Campaigns and accounting.}
The target controls in Table~\ref{tab:webshop-transfer-analysis}, Panel B, rescore saved trajectories from five primary cold-start \webshop{} campaigns per guidance condition, without a source constraint.
Each campaign evaluates the initial program and 40 generated candidates on the same 500 adaptation tasks: 41 candidate evaluations and 20,500 trajectory records per campaign.
All scoring judges receive identical reward-scrubbed trajectories within each campaign; agent execution is held fixed.
Original predictions are retained when available, with fresh predictions obtained for cross-scoring.
Target MAE is computed within each campaign and then averaged over its guidance condition's campaigns; source MAE is computed on the fixed source holdout.

\paragraph{Final selection and reported values.}
Within each campaign, each judge selects the candidate with the highest mean predicted score across the 500 tasks; ties favor the earlier candidate.
Audited selected reward, oracle-best reward, and regret are computed within the campaign before averaging.
Table~\ref{tab:webshop-transfer-analysis}, Panel B, is the main report of cross-scoring prediction and selection results.
Primary selected rewards and within-campaign regret are reported in Table~\ref{tab:main-results}; the final-selection accounting in Section~5.3 uses those displayed means.

\paragraph{Source-fitted isotonic calibration.}
Let $s_i$ be the fixed-rubric judge's score and $r_i$ the source evaluator reward for training trajectory $i$.
We fit a non-decreasing mapping by unweighted least-squares isotonic regression \citep{zadrozny2002transforming}:
\[
\hat g \in \operatorname*{arg\,min}_{g\ \mathrm{nondecreasing}}
\sum_{i=1}^{500}\bigl(g(s_i)-r_i\bigr)^2.
\]
Predictions outside the observed training-score range use the fitted boundary values.
The calibrated fixed-rubric judge returns $\hat g(s)$ without changing trajectory interpretation or diagnostics.
No validation, holdout, or target labels are used to fit the map or select its settings.
Although the mapping preserves trajectory-level score order up to ties, applying a nonlinear transformation before averaging can change candidate rankings.
The fitting objective is squared error, whereas source judge evolution optimizes MAE.

\paragraph{Deterministic scoring.}
The deterministic evolved judge extracts trajectory features, computes a structural score, and applies the evolved postprocessing rules with LLM-generated fields left empty.
It retains the extraction rules, scoring tables, calibration functions, and missing-field defaults, and obtains no fresh LLM judgment.
LLMs are still used during judge development and to generate the agent trajectories.
This intervention uses the frozen evolved judge selected by source-validation MAE for the primary \webshop{} experiments.
Comparing its code with the seed judge confirms that the independent numerical scoring pathway appeared during source evolution.
LLM-dependent inputs are removed together.

\paragraph{Offline prefix selection.}
For campaign $c$, let $\mathcal P_{c,t}$ contain its candidates generated through iteration $t$.
Define
\[
B_{c,t}=\max_{\pi\in\mathcal P_{c,t}}\bar R_c(\pi),\qquad
S_{c,t}(J)=\bar R_c\!\left(\operatorname*{arg\,max}_{\pi\in\mathcal P_{c,t}}\bar J_c(\pi)\right),
\]
where bars denote means over the fixed adaptation tasks and ties follow the earlier-candidate rule.
Prefix regret is $B_{c,t}-S_{c,t}(J)$.
Within each campaign, we count positive-regret prefixes among its 40 post-initialization stages and compute its maximum prefix regret; each statistic is averaged across five campaigns.

The main paper reports the mean number of prefixes with positive regret (out of 40 per campaign) and the mean of the campaign-specific maximum regrets.
These comparisons describe prediction and selection on saved candidates; they do not isolate the effects of promotion, diagnostics, and proposal randomness on candidate generation.

\FloatBarrier

\section{Judge Baselines and Component Ablation Results}
\label{app:online-ablation-results}

All conditions use five independent target-adaptation campaigns starting from the same unevolved agent program (cold start). We retain the primary WebShop task sets: 500 clothing source-training tasks, 300 source-validation tasks, 300 source-holdout tasks, 500 non-clothing adaptation tasks, and 500 disjoint held-out target tasks. Prompt-only optimization and executable evolution start from the same judge implementation and use identical source trajectories, validation procedures, and budgets of 20 proposer rounds. Training MAE determines the evolutionary parent, and private source-validation MAE selects the final judge; source-holdout data are excluded from development. The selected judges remain frozen during target adaptation. Each campaign uses 40 search rounds, and at most 100 environment steps per trajectory. This separate ablation series retains the primary data and search protocol while varying the judge-development procedure and the availability of diagnostics or numerical score adjustments.

Table~\ref{tab:online-evaluator-ablations} gives the numerical results plotted in Figure~\ref{fig:online-evaluator-ablations}.
Values are mean true rewards and standard deviations across campaigns for each condition.
Held-out evaluation uses frozen selected programs on separate target tasks.

\begin{table}[H]
\centering
\caption{Judge baselines and component ablations. Mean true reward $\pm$ standard deviation across campaigns for each condition.}
\label{tab:online-evaluator-ablations}
\begin{tabularx}{\linewidth}{@{}Xrr@{}}
\toprule
Judge and feedback condition & Adaptation & Held-out \\
\midrule
Evolved judge: scores + diagnostics & $\mathbf{0.6984}\pm0.0730$ & $\mathbf{0.6869}\pm0.0613$ \\
Unchanged seed judge & $0.2805\pm0.1664$ & $0.3329\pm0.1811$ \\
Source-optimized prompt-only judge & $0.5511\pm0.0591$ & $0.5795\pm0.0417$ \\
Evolved judge: scalar feedback only & $0.6579\pm0.0216$ & $0.6608\pm0.0175$ \\
Evolved judge without score adjustments: scalar feedback only & $0.6368\pm0.0362$ & $0.6124\pm0.0311$ \\
\bottomrule
\end{tabularx}
\end{table}

\end{document}